\documentclass[aps,pre,twocolumn,float]{revtex4}
 \usepackage{amssymb,amsmath,bm,epsfig}
\def\Fbox#1{\vskip1ex\hbox to 8.5cm{\hfil\fboxsep0.3cm\fbox{%
  \parbox{8.0cm}{#1}}\hfil}\vskip1ex\noindent}  %%  {TEXT} in BOX
\let \nn  \nonumber

\let\*\cdot
\def\<{\left\langle} \def\>{\right\rangle} \def\({\left(} \def\){\right)}
\let\p\partial \let\~\widetilde \let\^\widehat 

\newcommand{\C}[1]{{\mathcal{#1}}}    %%   Calligrapfic Upper case

 \newcommand{\BE}[1]{\begin{equation}\label{#1}}
 \newcommand{\BEA}[1]{\begin{eqnarray}\label{#1}}
 \newcommand{\BSE}[1]{\begin{subequations}\label{#1}}
 \def\bc{\begin{cases}}\def\ec{\end{cases}}
 \def\bse{\begin{subequations}}
 \def\ESE{\end{subequations}}

\newcommand{\eq}[1]{(\ref{#1})}%%  requires \eq{label}
\newcommand{\Eq}[1]{Eq.~(\ref{#1})}%%  requires \eq{label}
\newcommand{\Eqs}[1]{Eqs.~(\ref{#1})}%%  requires \eq{label}
\newcommand{\Ref}[1]{Ref.~\cite{#1}}%%  requires \Fef{label}
\let \nn  \nonumber  \let \= \equiv

\def\p{\partial}

\def\pz{\partial_z}

\begin{document}

\title{Symmetries and Interaction coefficients of Kelvin waves}

\author{Vladimir V. Lebedev$^\dag$ and  Victor S. L'vov$^{*}$}

 \affiliation{$^\dag$ Landau Institute for theoretical physics RAS,
 Moscow, Kosygina 2, 119334, Russia\\
 $^*$ Department of Chemical Physics, The Weizmann Institute of Science, Rehovot 76100, Israel}

   \begin{abstract}
We considered symmetry restriction on the interaction coefficients of Kelvin waves and
demonstrated that linear in small wave vector asymptotic is not forbidden, as one can
expect by na\"{\i}ve reasoning.
    \end{abstract}

\pacs{47.27.-i, 47.10.+g, 47.27.Gs}

\maketitle

\section*{Introduction}

Kelvin-wave energy cascade is believed to be a relevant ingredient of the quantum
turbulence realized at low temperatures in superfluids \cite{Vinen}. The cascade
transfers energy from the intervortex distance to smallest scales determined by the
vortex core radius. A theory of the cascade can be constructed in spirit of weak
turbulence theory \cite{ZLF} starting from the Hamiltonian representation of the
equations describing the vortex dynamics \cite{Svistunov}. Due to the one-dimensional
nature of the Kelvin waves their six-wave interaction is relevant for the cascade
\cite{Kozik,KS-rev}. Therefore a central question concerning the local character of the
cascade is related to asymptotic behavior of the corresponding interaction vertices of
the Kelvin waves.

Recent discussion \cite{Kozik,KS-rev,LLNR,LvovNaz} on the local/nonlocal nature of the
Kelvin wave (KW) cascade raised an important question concerning asymptotical behavior of
six-KW interaction amplitude
 \begin{subequations}\label{res1}
 \begin{equation}\label{res1A} \C   W_{1,2,3}^{4,5,6}
   \equiv  \C  W(k_1,k_2,k_3|k_4,k_5,k_6)
 \end{equation}
in the  interaction Hamiltonian
 \begin{equation}\label{res1B}
 \C H_6= \frac {1}{36} \sum_{k_j}
 \C   W_{1,2,3}^{4,5,6}w_1w_2w_3w_4^*w_5^*w_6^* \Delta _{1,2,3}^{4,5,6}\ .
 \end{equation}\end{subequations}
Here $w_j\=w(k_j)$ is  an   amplitude of KWs with wave vector $k$, a Fourier transform of
a (complex) two-dimensional displacement vector $w(z)=x+iy$ of a vortex line from the
straight line $x=y=0$. The summation over six $k$-vectors $k_j=k_1\dots k_6$ is
restricted by the Kroneker symbol, $ \Delta _{1,2,3}^{4,5,6}$, equal to one if
$k_1+k_2+k_3=k_4+k_5+k_6$, and to zero otherwise.

Explicit (and cumbersome) calculations, see \Ref{LLNR}, gave the value of $\C
W_{1,2,3}^{4,5,6} $  in the asymptotical limit, when one or few wave vectors are much
smaller than others:
 \begin{subequations}\label{res2}
 \begin{equation}\label{res2A}
 \C  W_{1,2,3}^{4,5,6}
   =-\frac3{4\, \pi}\,   k_1 k_2 k_3 k_4 k_5 k_6  \ .
\end{equation}
It is important that \Eq{res2A} gives
 \begin{equation}\label{res2B}
 \C W\propto k_1\,, \quad \mbox{for} \quad
 k_1\ll k_0\sim  k_2 \sim k_3\sim \dots   k_6\ .
 \end{equation}
If so, then the interaction vertex of short-wave motions (with wave vectors $k_j\sim
k_0$) with long-wave ``$k_1$-motions" is proportional to $k_1  w_1$, or, in the physical
space, is proportional to the spatial $z$-derivative of the displacement, $\partial_z
w(z)$.

One may think \cite{K-Lammi} that this asymptotic behavior contradicts to physical
intuition, according to which the interaction cannot depend on the local slope of
long-wave disturbances, $\partial_z w(z)$, because of rotational invariance of the
interaction Hamiltonian $\C H_6$. Indeed, one can choose the coordinate system, oriented
along the (local) direction of the long-wave disturbances, in which $\partial_z w(z)=0$.
Then the curvature of the long-wave disturbances (proportional to the second derivative
$\p^2_z w(z)=0$) is expected \cite{K-Lammi}  to be relevant. If this is true, then
instead of asymptote~\eq{res2B} one has
 \begin{equation}\label{res2C}
 \C W\propto k_1^2\,, \quad \mbox{for} \quad   k_1\ll k_0
 \simeq  k_2 \simeq  k_3\simeq \dots   k_6\ .
 \end{equation}
 \end{subequations}
So, the dilemma is: either the cumbersome calculations \cite{LLNR} are mistaken and
linear asymptote~\eq{res2B} is wrong, or something is wrong with the simple symmetry
analysis \cite{K-Lammi}, leading to quadratic asymptote~\eq{res2C}.

The difference between the asymptote ~\eq{res2B} and the asymptote \eq{res2C} is of
crucial importance for the physics of KW energy cascade:
\begin{itemize}

 \item
In the case of  quadratic asymptote~\eq{res2C} the energy cascade should  be dominated by
local, step-by step energy transfer by interacting KWs with wave vectors of the same
order of magnitude $k_1\sim k_2\sim k_3\sim k_4 \sim k_5 \sim k_6$. This scenario leads
to the Kosik-Svistunov energy spectrum \cite{Kozik} with
 \begin{equation}
 E(k)\propto k^{-17/5}.
 \label{KS}
 \end{equation}

 \item
In the case of the linear asymptote~\eq{res2B} the interactions between KWs in sixtets
with very different wave vectors are much stronger and are dominated by  the region,
where two of three wave vectors from the sextet are of the order of inverse intervortex
distance, $1/\ell$, and much smaller than the other four ones. For example, $k_1\sim k_2
\sim   k_3\sim k_4 \gg   k_5 \sim k_6  \sim \frac 1 \ell$. In this case the curvature of
vortex lines of intervortex scales opens new ``quartet" channel of an effective four-wave
interaction with $ k_1+ k_2 +   k_3= k_4$, that leads to the energy spectrum recently
found by L'vov and Nazarenko \cite{LvovNaz}
 \begin{equation}
 E(k)\propto k^{-5/3}.
 \label{LN}
 \end{equation}

\end{itemize}

\section*{I.~Rotational symmetry and line length}

To shed light on the contradiction between the asymptotics  \eq{res2B} and \eq{res2C} let us
consider a simple object: the length $L$ of a self-affine (without overhangs) line
described by $x+iy=w(z)$ and fixed at the points $x=y=z=0$ and  $x=y=0,\ z=L_0$. This
length is given by the formally exact expression
  \begin{equation}\label{res3}
 L=\int\limits _0^{L_0}\sqrt{1+|\partial_z w(z)|^2}dz\ .
 \end{equation}
For small tilt, $|\partial_z w(z)|\ll 1$, the expression \Eq{res3} can be expanded as
follows:
 \begin{subequations}\label{res4}
 \begin{equation}\label{res4a}
 L=L_0+ L_2 + L_4 + L_6+\dots\,,
 \end{equation}
where
 \begin{equation}\label{res4b}
 L_2=\frac12 \int\limits _0^{L_0} |\partial_z w(z)|^2\,
 dz= \frac12 \sum_k k^2 |w_k|^2 \,,
 \end{equation}\end{subequations}
describes individual contributions of waves with different wave vectors $k$ to the line
length.

Higher terms are responsible for the cross-contributions to $L$ from waves with different
$k$. For example:
 \begin{subequations}\label{res5}
 \begin{eqnarray}\label{res5A}
 L_4&=&-\frac 18 \int\limits _0^{L_0} |\partial_z w(z)|^4\,  dz
 \\ \nn &=&
  \sum_{k_i} T_{1,2}^{3,4}  w_1 w_2 w_3^* w_4^* \Delta_{1,2}^{3,4}\,;
  \\ \label{res5B}
 && T_{1,2}^{3,4}=-\frac 18 \, k_1 k_2 k_3 k_4\,, \\
 \label{res5C} L_6&=&\frac 1{16} \int\limits _0^{L_0} |\partial_z w(z)|^6\,  dz
 \\ \nn  &=&
 \sum_{k_j}W_{1,2,3}^{4,5,6} w_1 w_2 w_3 w_4^*w_5^*w_6^* \Delta_{1,2,3}^{4,5,6} \,,
 \\ \label{res5D} &&
 W_{1,2,3}^{4,5,6} = \frac1{16}\,   k_1 k_2 k_3 k_4 k_5k_6 \ .
 \end{eqnarray}
 \end{subequations}
As one sees from \Eq{res5D}, the vertices $T_{1,2}^{3,4}$ and $W_{1,2,3}^{4,5,6}$ have
exactly the same linear in $k$ dependence as $\C W_{1,2,3}^{4,5,6}$ in \Eq{res2A} and
thus have the same problems with the ``na\"{\i}ve" physical intuition, reproduced above.
Indeed, repeating the same reasoning, one may think \cite{K-Lammi} that effect of the
long-wave disturbances cannot depend on its local slope, because ``one can choose the
coordinate system, oriented along the (local) direction of the long-wave disturbances, in
which $\partial_z w(z)=0$". If so, the vertices $T_{1,2}^{3,4}$ and $W_{1,2,3}^{4,5,6}$
have to be proportional to the square of the wave vector $k_1$ of the long-wave
disturbances, and not to its first power, as in \Eq{res5D}. However, the expressions
\Eqs{res5} are definitely correct, being simple straightforward consequence of the
expression \Eq{res3}.

To resolve this ``contradiction" we will elaborate some consequences of the rotational
symmetry of \Eq{res3} for the line length. For this purpose we introduce ``slow" line
displacement $\xi(z)=x+iy$ [in the original, global $(x,\, y,\, z)$-reference system] and
``fast" line displacement $u(z)=\tilde  x + i \tilde  y$ in the ``local" ($\tilde x,\,
\tilde y,\, \tilde z$)-reference system with the origin following the slow displacement
$\xi (z)$ and $\tilde  z$ axis oriented along the local direction of the slow line $\xi
(z)$. Then for small slow slopes, $\partial_z \xi (z)\ll 1$, the total line displacement
in the global reference system can be approximated as follows:
 \begin{equation}
 w(z) =\xi(z) + u(z) + \mbox{Re}[\partial_z \xi(z)  u^*(z)] \partial_z  u(z)\ .
 \label{decompos}
 \end{equation}
The last term here originates from the rotation of $\tilde z$-axis from the original
direction of $z$-axis.

Now, in the ($\tilde x,\, \tilde y,\, \tilde z$)-reference system, we can compute
$\delta_{\xi  u} L$, the cross-contribution to the line length caused by combine effect
of the slow and fast displacements. Substituting $w(z)$ from \Eq{decompos} into \Eq{res3}
and integrating by parts, one gets in the linear in $\xi $ approximation:
 \begin{equation}
 \delta_{\xi  u}   L =\int dz\, \mbox{Re}[\Phi^* \partial_z^2\xi(z)]\,,
 \  \Phi \= \frac { u(z) }
 {\sqrt{1+|\partial_z   u(z)|^2}}\ .
 \label{deltal}
 \end{equation}
In agreement with the symmetry reasoning the variation (\ref{deltal}) is proportional to
$\partial_z^2\xi$ that is determined by the curvature of the line $w=\xi$.

However the line length~\eq{res3} and its expansion~\eq{res5} are written in terms of
$w(z)$, (i.e. in the global reference system) while the variation~\eq{deltal} is
presented via $\xi(z)$ in the local reference system. To rewrite the result~\eq{deltal}
in terms of $ w(z)$, one should use a transformation, inverse to Eq. (\ref{decompos}). It
can be done by iterations. The zero-order term is
 \begin{subequations} \label{iter}
 \begin{equation}
 \xi_0(z)=w(z)-u(z)\ ,
 \label{iter0}
 \end{equation}
and the first iteration is
 \begin{equation}
 \xi_1(z)=\xi_0(z) -\mbox{Re} [\partial_z\xi_0(z) \, u^*(z)] \partial_z   u(z)\,,
 \label{iter1}
 \end{equation}\end{subequations}
where $u(z)$ can be substituted by the fast part of $w(z)$. Substituting \Eq{iter} into
\Eq{deltal} one finds:
 \begin{eqnarray}\label{corr}
  \delta L&=&-\int dz\, \mbox{Re}\Big \{\Phi^*  \Big [  \pz \xi_0(z) \p^2 _z
  \big [ u^* (z) \pz u(z)\big ]\\ \nn  &&
  +   \pz \xi_0(z) \p^2 _z \big [ u (z) \pz u^* (z)\big ]  \Big] \Big \}+ \dots \ .
 \end{eqnarray}
This equation explicitly contains the first derivative of the slow displacement
$\xi_0(z)$. Therefore the rotational symmetry does not forbid linear in small $k_1$ terms
in the expansion \eq{res4a}. It is confirmed by simple calculations leading to the
expressions (\ref{res5}).

\section*{II. Long-scale behavior of interaction vertices}

Here we show that the rotational symmetry also does not contradict to the linear in $k_1$
asymptote~\eq{res2B} of the vertex $\C  W_{1,2,3}^{4,5,6}$, \Eq{res2A}. To see this we
consider   a single quantum self-affine vortex of length $L$, given by \Eq{res3} and
fixed, as before, at the boundary points. At zero temperature the vortex dynamics is
determined by the formally exact Hamiltonian
 \begin{eqnarray}
 {\cal H}&=&\frac{\kappa}{4\pi}
 \int \frac{dz_1\, dz_2} R \, \Big\{ 1+\mbox{Re}
 \Big[\frac{\partial  w_1 }{\partial z_1}
 \frac{\partial w_2^*}{\partial z_2}  \Big]\Big\}
 \label{hamil} \\
 &=&-\frac{\kappa}{\pi}\int dz_1\, dz_2\,
 \frac{\partial}{\partial z_1} \sqrt R\,
 \frac{\partial}{\partial z_2} \sqrt R,
 \nonumber
 \end{eqnarray}
with
 \begin{equation}
 R^2 =|w(z_1)-w(z_2)|^2+(z_1-z_2)^2 +a^2,
 \label{separ}
 \end{equation}
suggested by Sonin \cite{Sonin} (see also Ref. \cite{Svistunov}). Here $\kappa$ is the
circulation quantum and $a$ is the vortex core size (introduced for regularization).

Let us repeat for the Hamiltonian (\ref{hamil}) the same logic steps as  for the length
$L$. Now we are interested in an expansion of the Hamiltonian over the slow variable
$\xi$ and define the fast variable $u$ in the reference system attached to the slow
variable. The first term of the expansion of the Hamiltonian can be found by a bit more
complicated calculations than the ones leading to \Eq{deltal}. First of all, we find
\begin{widetext}
 \begin{eqnarray}
 \delta R^2 &=&\mbox{Re}\Big\{
 ( \xi_1'  u_1^*) \frac{\partial}{\partial z_1} R^2
 +( \xi_2'  u_2^*) \frac{\partial}{\partial z_2} R^2
  +2 u_1^* [ \xi_1- \xi_2- \xi_1'(z_1-z_2)]
  +2 u_2^* [ \xi_2- \xi_1- \xi_2'(z_2-z_1)]\Big\}
 \nonumber \\
 &\to& \mbox{Re}\Big\{ ( \xi_1'  u_1^*) \frac{\partial}{\partial z_1} R^2
 +( \xi_2'  u_2^*) \frac{\partial}{\partial z_2} R^2
  +(z_1-z_2)^2 ( u_1^* \xi_2 ''+  u_2^*  \xi_1 '')\Big\}  \,,
 \label{interm}
 \end{eqnarray}
where prime means derivative over $z$. The transformations of the expressions in square
brackets are justified since $\xi$ is the soft variable. Next, the variation of the
Hamiltonian (\ref{hamil}) can be written as
 \begin{eqnarray} \label{varham}
 \delta _{\xi  u}  {\cal H} &=&-\frac{2\kappa}{\pi}
 \int dz_1\, dz_2\, \frac{\partial}{\partial z_1}\sqrt R\,
 \frac{\partial}{\partial z _2}\delta \sqrt R
 \nonumber \\
 &\to&  -\frac{2\kappa}{\pi}
 \int dz_1\, dz_2\, \frac{\partial}{\partial z _2}\left[
 (\bm \xi_2' \bm u_2)\frac{\partial}{\partial z_1} \sqrt R
 \frac{\partial}{\partial z_2} \sqrt R\right]
 \\ \nonumber
 && -\frac{\kappa}{2\pi}
 \int dz_1\, dz_2\, \frac{\partial}{\partial z_1}\sqrt R\,
 \frac{\partial}{\partial z_2}\left[
 R^{-3/2}(z_1-z_2)^2 \mbox{Re}\left( u_1^*   \frac{\partial^2 \xi_2 }{\partial z_2^2}+
 u_2^* \frac{\partial^2 \xi_1 }{\partial z_1^2}\right) \right] \ .
 \end{eqnarray}
 \end{widetext}
The second line in Eq. (\ref{varham}) disappears after integration in part. Thus we
conclude that the principal contribution to the first-order term of the Hamiltonian
expansion is proportional to $\partial_z^2 \xi$, i.e. can be written in the form similar
to Eq. (\ref{deltal}):
 \begin{equation}
 \delta \C H =\int dz\, \mbox{Re}[\Psi^* (z)\partial_z^2\xi(z)]\ .
 \label{deltaH}
 \end{equation}
An explicit expression for $\delta \Psi(z)$ can be found from Eq. (\ref{varham}). The
expression (\ref{deltaH}) is in accordance with the symmetry expectations.

Now, as before, we should return to original variables substituting $\xi(z)$ from
\Eq{iter}. The resulting expression for $\delta \C H$ can be obtained from \Eq{corr} by
replacing $\Phi\to \Psi$. Therefore the expression for $\delta \C H$ contains the first
derivative of the slow variable $\xi_0$. Thus the interaction amplitude  $\C
W(k_1,k_2,k_3|k_4,k_5,k_6)$  with the linear in the wave vector $k$ long-scale
asymptote~\eq{res2A} is not forbidden by the rotational symmetry.

\section*{Conclusion}

We found that the linear in small wave vector asymptote~\eq{res2B} of the interaction
vertices of Kelvin waves $\C  W(k_1,k_2,k_3|k_4,k_5,k_6)$~\eq{res2A}, which  results in
the LN energy spectrum~\eq{LN}, is not forbidden by the rotational symmetry.

\acknowledgments

These  notes have become possible due to kind hospitality at Lammi (Finland) Symposia On
Superfluids Under Rotation, 11-16 April 2010, prolonged by the Eyjafjallajokull volcano
eruption. Useful discussions with Sergei Nasarenko are highly appreciated.

\end{document}